\documentclass[aps,twocolumn,prl,showpacs,amsmath]{revtex4}

\usepackage{graphicx}
\usepackage{bm}

\newcommand{\figurewidth}{3.4in}

\begin{document}

\title{Screening in Ionic Systems: Simulations for the Lebowitz Length}

\author{Young C. Kim}
\author{Erik Luijten}
\altaffiliation{Now at: Department of Materials Science and Engineering, University of Illinois, Urbana, Illinois 61801, USA}
\author{Michael E. Fisher}
\email[Corresponding author: ]{xpectnil@ipst.umd.edu}
\affiliation{Institute for Physical Science and Technology, University of Maryland, College Park, Maryland 20742, USA}

\date{\today}

\begin{abstract}
Simulations of the Lebowitz length, $\xi_{\text{L}}(T,\rho)$, are reported for the restricted primitive model hard-core (diameter $a$) 1:1 electrolyte for densities $\rho\lesssim 4\rho_c$ and $T_c \lesssim T \lesssim 40T_c$. Finite-size effects are elucidated for the charge fluctuations in various subdomains that serve to evaluate $\xi_{\text{L}}$. On extrapolation to the bulk limit for $T\gtrsim 10T_c$ the low-density expansions (Bekiranov and Fisher, 1998) are seen to fail badly when $\rho > \frac{1}{10}\rho_c$ (with $\rho_c a^3 \simeq 0.08$). At higher densities $\xi_{\text{L}}$ rises above the Debye length, $\xi_{\text{D}} \propto \sqrt{T/\rho}$, by 10-30$\%$ (upto $\rho\simeq 1.3\rho_c$); the variation is portrayed fairly well by generalized Debye-H\"{u}ckel theory (Lee and Fisher, 1996). On approaching criticality at fixed $\rho$ or fixed $T$, $\xi_{\text{L}}(T,\rho)$ remains finite with $\xi_{\text{L}}^c \simeq 0.30 a \simeq 1.3 \xi_{\text{D}}^c$ but displays a weak entropy-like singularity.

\end{abstract}
\maketitle

Understanding the thermodynamic and correlation properties of ionic fluids has challenged both theory and experiment \cite{wei:sch}. Typical electrolytes exhibit phase separation that is analogous to the gas-liquid transition in simple fluids, albeit at rather low temperatures when appropriately normalized. However, the long range of the Coulomb interactions has hampered understanding especially near criticality \cite{wei:sch}. One crucial aspect is Debye-H\"{u}ckel screening. For a $d$-dimensional classical fluid system with short-range ion-ion potentials beyond the Coulomb coupling $z_\sigma z_\tau q^2/r^{d-2}$ (where $z_\sigma$ is the valence of ions of species $\sigma$ and mole fraction $x_\sigma$ while $q$ is an elementary charge), the charge-charge correlation function, $G_{ZZ}(\mbox{\boldmath $r$};T,\rho)$, decays as $\exp[-|\mbox{\boldmath $r$}|/\xi_{Z,\infty}(T,\rho)]$ (see, e.g., \cite{bek:fis,aqu:fis}): the asymptotic screening length, $\xi_{Z,\infty}$, approaches the Debye length $\xi_{\text{D}} {\,=\,} (k_{\text{B}}T/4\pi \bar{z}_2^2 q^2\rho)^{1/2}$ when the overall ion density $\rho$ approaches zero (with $\bar{z}_2^2 {\,=\,} \sum_\sigma z_\sigma^2 x_\sigma$ \cite{bek:fis,aqu:fis}).

By contrast, at a critical point of fluid phase separation, the density-density (or composition) correlation length, $\xi_{N,\infty}(T,\rho)$, {\em diverges}, as do all the {\em moments} of $G_{NN}(\mbox{\boldmath $r$};T,\rho)$. What then happens to charge screening near criticality? This question was first posed over a decade ago \cite{ste} and has been addressed recently via the exact solution of $(d {\, >\,} 2)$-dimensional ionic spherical models \cite{aqu:fis}. As anticipated [4(b)], the issue of $\pm$ {\em ion symmetry} proves central. However, spherical models for fluids display several artificial features (e.g., infinite compressibilities on the phase boundary below $T_c$; parabolic coexistence curves, $\beta\equiv\frac{1}{2}$; etc.). Accordingly, understanding screening near criticality for more realistic models remains a significant task.

To that end we report here on a Monte Carlo study of the restricted primitive model (RPM), namely, hard spheres of diameter $a$ carrying charges $q_\pm = \pm q$ (so that $z_+ = -z_- = 1$, $x_+ = x_- = \frac{1}{2}$). Grand canonical simulations have been used and, to accelerate the computations, a finely discretized ($\zeta {\,=\,} 5$ level) lattice version of the RPM has been adopted \cite{pan}. For this system the critical behavior is well established as of Ising-type with $T_c^\ast {\,\equiv\,} k_{\text{B}} T_c a/q^2 {\,\simeq\,} 0.05069$ and $\rho_c^\ast \equiv \rho_c a^3 \simeq 0.079$ \cite{lui:fis:pan}. Furthermore, it has been demonstrated that for $\zeta \gtrsim 3$ the fine-lattice discretization does not qualitatively affect thermodynamic or finite-size properties \cite{kim:fis}.

Ideally one would like to calculate $\xi_{N,\infty}(T,\rho)$ and $\xi_{Z,\infty}(T,\rho)$ near criticality; but even in {\em non}ionic model fluids, obtaining $\xi_{N,\infty}$ via simulations is hardly feasible. Nevertheless, the low-order moments $M_{N,k} = \int |\text{\boldmath $r$}|^k G_{NN} (\text{\boldmath $r$}) d^d r$ for $k=0,1,2,\cdots$, are accessible and, by scaling, all the $\xi_{N,k}\equiv (M_{N,k}/M_{N,0})^{1/k}$ for $k > 0$ diverge like $\xi_{N,\infty}$. However, for charges the Stillinger-Lovett sum rules \cite{bek:fis,aqu:fis} dictate $M_{Z,0} \equiv 0$ (so that $G_{ZZ}(\text{\boldmath $r$})$ is not of uniform sign) while the second moment satisfies $M_{Z,2} = 2\bar{z}_2^2 q^2 \rho \xi_{\text{D}}^2$ which is fully {\em analytic} through $(T_c,\rho_c)$. On the other hand, the {\em first moment} of $G_{ZZ}(\text{\boldmath $r$})$ is known \cite{bei:fel} to be intimately related to charge screening via the so-called ``area law'' of charge fluctuations.

To explain this, consider a regular subdomain $\Lambda$ with surface area $A_\Lambda$ and volume $|\Lambda|$, embedded in a larger domain, specifically say, the cubical $L^d$ simulation box. If $Q_\Lambda$ is the total fluctuating charge in $\Lambda$, electroneutrality implies $\langle Q_\Lambda \rangle = 0$; but the mean square fluctuation, $\langle Q_\Lambda^2 \rangle$, will grow when $|\Lambda|$ increases. In the {\em absence} of screening one expects $\langle Q_\Lambda^2 \rangle \sim |\Lambda|$; however, in a fully screened, bulk $(L\rightarrow\infty)$ conducting fluid $\langle Q_\Lambda^2 \rangle$ is asymptotically proportional to the surface area \cite{bei:fel}. This was first observed by van Beijeren and Felderhof and later proven rigorously by Martin and Yalcin \cite{bei:fel}. Following Lebowitz \cite{bei:fel} one may then define a screening distance proportional to $M_{Z,1}(T,\rho)$, which we call the {\em Lebowitz length}, $\xi_{\text{L}}(T,\rho)$ \cite{bek:fis} via
\begin{equation}
  \langle Q_\Lambda^2 \rangle/A_\Lambda \approx c_d \rho \bar{z}_2^2 q^2 \xi_{\text{L}}(T,\rho) \hspace{0.3in} \mbox{as} \hspace{0.3in} |\Lambda|\rightarrow\infty,  \label{def.leb}
\end{equation}
where $c_d$ is a numerical constant with $c_3 =\frac{1}{2}$. Note that, since $G_{ZZ}(\text{\boldmath $r$})$ is not necessarily of uniform sign, $\xi_{\text{L}}(T,\rho) \propto M_{Z,1}(T,\rho)$ might diverge at $T_c$ even though the second moment $M_{Z,2} \propto \xi_{\text{D}}^2$ remains finite!

Clearly, by simulating $\langle Q_\Lambda^2 \rangle$ in various subdomains one may, as we show here, hope to calculate the Lebowitz length. To our knowledge no numerical results have been reported previously for $d=3$ although Levesque {\em et al.} \cite{lev:wei:leb} presented a study (above criticality) for $d=2$. An exact low density expansion \cite{bek:fis} proves that $\xi_{\text{L}}/\xi_{\text{D}}\rightarrow 1$ when $\rho\rightarrow 0$ and corrections of order $\rho^{1/2}$, $\rho\ln\rho$ and $\rho$ have been evaluated. This analysis \cite{bek:fis} also served to validate the {\em generalized Debye-H\"{u}ckel} (GDH) theory for the correlations \cite{lee:fis} for {\em small} $\rho$.

The GDH theory, however, did not generate a $\rho\ln\rho$ term: nevertheless, as we find here, the exact expansion fails at very low densities --- around $\rho_c/10$ even for $T\simeq 10 T_c$ --- while GDH theory provides a reasonable estimate of $\xi_{\text{L}}(T,\rho)$ at higher densities: see Fig.\ \ref{fig3} below. Furthermore, our calculations show that $\xi_{\text{L}}$ remains {\em finite} at criticality, exceeding $\xi_{\text{D}}^c$ by only $33\%$. Nonetheless, the Lebowitz length does exhibit {\em weak singular behavior} that, in accord with general theory, matches that of the entropy.

The first serious computational task is to understand the finite-size effects resulting from the $L\times L\times L$ simulation box with periodic
boundary conditions. Each simulation at a given $(T^\ast,\rho^\ast)$ yields a histogram of the total fluctuating charge $Q_{\Lambda}$ for 24 different subdomains~$\Lambda$. We have used: six small {\em cubes} of edges $\lambda L$ with $\lambda = 0.3,\, 0.4,\, \cdots,\, 0.8$; seven `{\em rods}' of dimensions $\lambda L {\,\times\,} \lambda L {\, \times\,} L$ with $\lambda=0.2,\,\ldots,\, 0.8$, four `{\em slabs}' of dimensions $\lambda L {\, \times\,} L {\, \times\,} L$ with $\lambda=0.2,\, \cdots,\, 0.5$; and seven {\em spheres} of radius $R=\lambda L$ with $\lambda = 0.15\,$-$\,0.45$ in increments $\Delta\lambda {\, = \,} 0.05$. To minimize correlations between these various subdomains, they have been located as far apart as feasible.

\begin{figure}[ht]
\centerline{\includegraphics[width=\figurewidth]{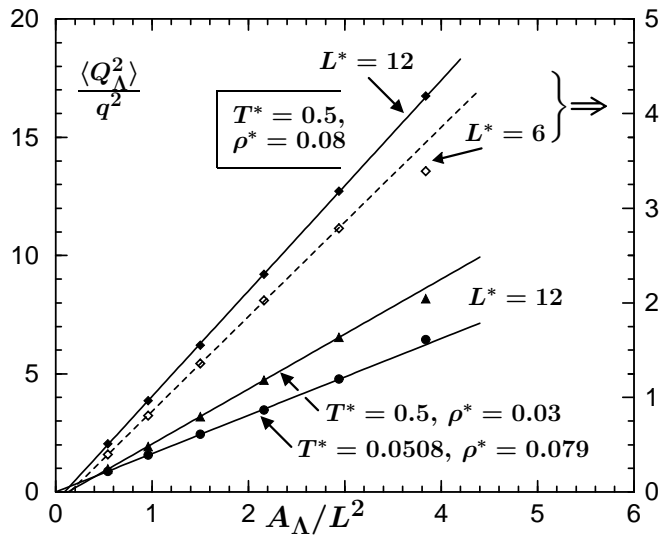}}
\caption{Reduced charge fluctuations, $\langle Q_\Lambda^2\rangle/q^2$, for given $(T,\rho)$ in cubes $\Lambda$ of edges $\lambda L$ vs.\ reduced area, $A_{\Lambda}/L^2=6\lambda^2$.}
\label{fig1}
\end{figure}

While the area law for the charge fluctuation, $\langle Q_{\Lambda}^2 \rangle$, is rigorously true for $L \rightarrow \infty$ followed by $\Lambda \rightarrow \infty$, it is by no means clear how it will be distorted for a finite subdomain $\Lambda$ embedded in a finite system.  To understand this Fig.~\ref{fig1} presents  $\langle Q_{\Lambda}^{2} \rangle$, normalized by $q^{2}$, for the six cubic subdomains as a function of the reduced area $A_{\Lambda}/L^2$ at selected temperatures and densities for box sizes $L^\ast \equiv L/a = 6$ and $12$. 
Surprisingly, at high temperature and moderate density ($T^\ast = 0.5 \simeq 10T_c^\ast,\rho^\ast = 0.08 \simeq \rho_c^\ast$), the area law is well satisfied for $\lambda \lesssim 0.7$ even for small systems. For $L^\ast = 6$ the data point for $\lambda = 0.8$ deviates strongly from the linear fit (dashed line) owing to finite-size effects: indeed, electroneutrality dictates that $\langle Q_\Lambda^2 \rangle$ should vanish when $\lambda \rightarrow 1$, corresponding to $A_\Lambda/L^2 = 6$. At low densities around $\frac{1}{3}\rho_c$, the Debye length $\xi_{\text{D}}\propto \sqrt{T/\rho}$ becomes large but nevertheless we see that the area law is still well satisfied. Furthermore, the area law is found to hold even near
criticality: see the lowest plot. Note, however, that the linear fits to the data do {\em not} pass through the origin. This reflects finite-size effects which are discussed further below.

Combining (\ref{def.leb}) with the observations illustrated in Fig.~\ref{fig1}, we conclude that charge fluctuations in the cubic subdomains are well
described by
\begin{equation}
  \langle Q_{\Lambda}^{2}(T,\rho;L) \rangle = A_{0}(T,\rho;L) + \tfrac{1}{2}\rho q^{2} \xi_{\text{L}}(T,\rho;L) A_{\Lambda} \;, \label{fit}
\end{equation}
where the intercept $A_{0}(T,\rho;L)$ need not vanish. The (fitted) linear slope serves to define the {\em finite-size Lebowitz length}, $\xi_{\text{L}}(T,\rho;L)$, which should approach the bulk value, $\xi_{\text{L}}(T,\rho)$. But by what route?

\begin{figure}[ht]
\centerline{\includegraphics[width=\figurewidth]{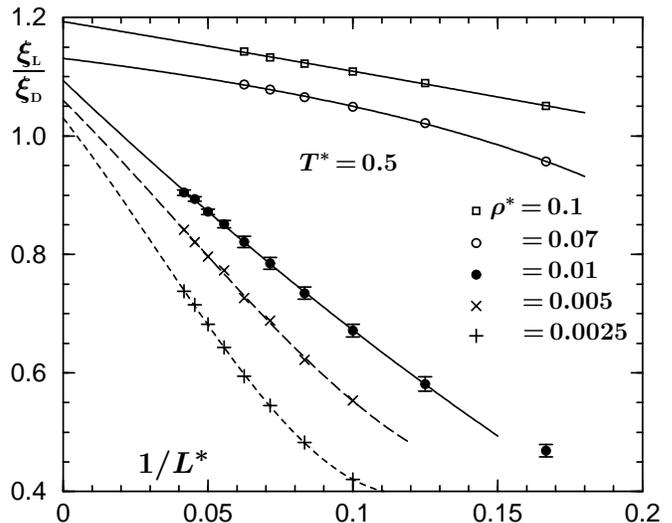}}
\caption{Quadratic fits to finite-size Lebowitz length data for sizes up to $L^{\ast}=24$ at $T^{\ast} = 0.5$ and various densities.}
\label{fig2}
\end{figure}

To answer this question consider Fig.~\ref{fig2} which displays $\xi_{\text{L}}(T,\rho;L)$ vs.\ $1/L^\ast$ for $T^\ast {\,=\,} 0.5$ at various densities. It is rather clear that $\xi_{\text{L}}(T,\rho;L)$ approaches its bulk limit as~$1/L$. This can be understood by recalling the Lebowitz picture \cite{bei:fel} in which the uncompensated charge fluctuations in a subdomain arise only from {\em shells} of area $A_\Lambda$ and thickness of order $\xi_{\text{L}}$. By invoking the screening of $G_{ZZ}(r)$ one can see that $\Delta\xi_{\text{L}} {\,\equiv\,} \xi_{\text{L}}(L) - \xi_{\text{L}}(\infty)$ for {\em smooth} subdomains decays as $1/L^2$. Indeed, by this route van Beijeren and Felderhof~\cite{bei:fel} showed explicitly that fluctuations in a sphere of radius $R$ (in an infinite system) approach their limiting behavior as $1/R^2$. For spheres in {\em finite} systems, we observe similarly that $\xi_{\text{L}}(L)$ approaches the bulk value as $1/L^2$. However, for cubes---which have edges and corners---and rods with edges, $\xi_{\text{L}}(L)$ gains a lower order, $1/L$ term as seen in Fig.~\ref{fig2}. (The intercept $A_0(L)$ in (\ref{fit}) is, correspondingly, found to vary as $L$.) On the other hand, for slabs, lacking edges and corners, we find that $\xi_{\text{L}}(L)$ obtained via (\ref{def.leb}) approaches the limit exponentially fast.

\begin{figure}[ht]
\centerline{\includegraphics[width=\figurewidth]{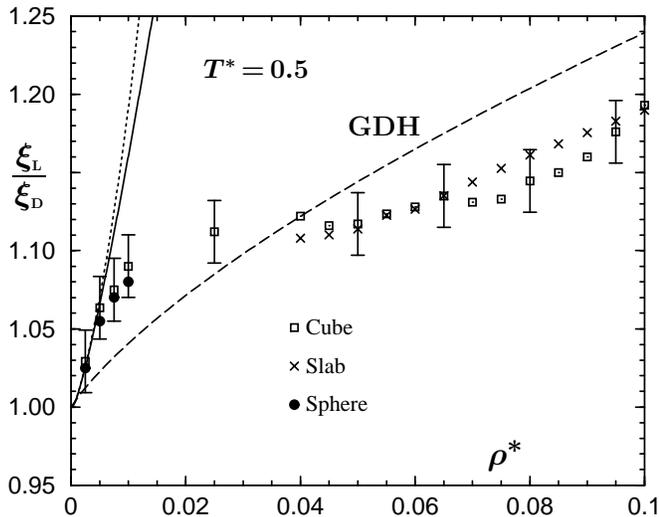}}
\caption{Density variation of the bulk Lebowitz length extrapolated from various subdomains at $T^\ast = 0.5$. The dashed, solid and dotted plots represent GDH theory~\cite{lee:fis}, and approximants exact at low density: see text.}
\label{fig3}
\end{figure}

Having established the finite-size behavior, let us examine $\xi_{\text{L}}(T,\rho)$ on the $T^\ast {\,=\,} 0.5$ isotherm, well above $T_c$. Figure~\ref{fig3} shows estimates extrapolated from cubes, spheres and slabs. At moderate densities systems up to $L^\ast {\,=\,} 16$ suffice but for $\rho^\ast {\,\leq\,} 0.025$ we went up to $L^\ast {\,=\,} 24$. The results may be compared with GDH theory \cite{lee:fis} (dashed curve) and approximants which reproduce the exact low-density expansion known to order $\rho$ \cite{bek:fis}. For the latter we adopt
\begin{eqnarray}
  \xi_{\text{L}}^{\text{[1,0]}} & = &
  \xi_{\text{D}}(T,\rho)\left [\, 1 + a_{1}(T)\rho^\ast + a_{2}(T)\rho^\ast \ln\rho^\ast
    \,\, \right ], \label{bf10}  \\
  \xi_{\text{L}}^{\text{[0,1]}} & = &
  \xi_{\text{D}}(T,\rho)/[1 - a_{1}(T)\rho^\ast - a_{2}(T)\rho^\ast \ln\rho^\ast],  \label{bf01}
\end{eqnarray}
shown in Fig.~\ref{fig3} as solid and dotted curves, respectively, where $a_1(T)$ and $a_2(T)$ follow from \cite{bek:fis}. The simulations agree well with the low-density expansion but only up to $\rho^\ast {\,\simeq\,} 0.005$; thereafter $\xi_{\text{L}}$ rises above the Debye length much more slowly. By contrast, GDH theory captures the overall behavior of $\xi_{\text{L}}(T,\rho)$ over a broad density range, representing the numerical estimates to within a few percent at moderate densities, $0.01 {\,\leq\,} \rho^\ast {\,\leq\,} 0.10$, where no exact results are available.

\begin{figure}[ht]
\centerline{\includegraphics[width=\figurewidth]{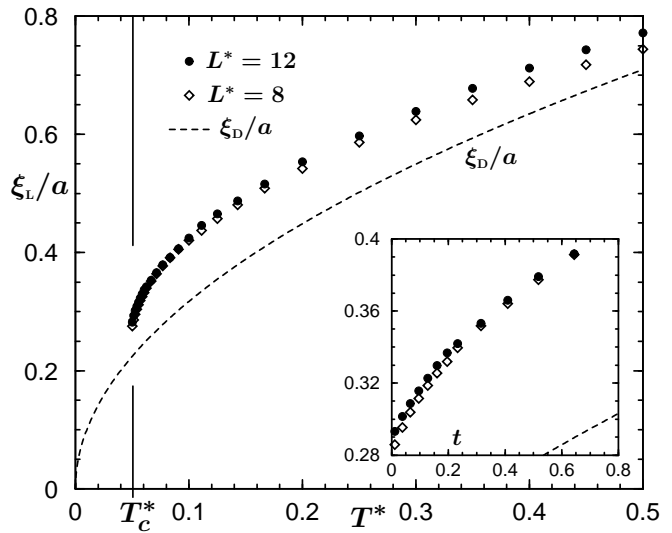}}
\caption{Lebowitz length for $L^\ast{\,=\,}8$ and 12 on the critical isochore compared with the Debye length.}
\label{fig4} 
\end{figure}

In the critical region the first question is the finiteness of $\xi_{\text{L}}(T_c,\rho_c)$. To answer we study $\xi_{\text{L}}$ on the critical isochore $\rho {\,=\,} \rho_c$ as $T {\,\rightarrow\,} T_c$. Figure~\ref{fig4} \cite{kim2} reveals that $\xi_{\text{L}}/a$ {\em falls} increasingly rapidly when $T^\ast$ drops from ${\sim\,} 0.5$ but clearly attains a finite nonzero value at $T_c$ that exceeds $\xi_{\text{D}}^c/a {\,\simeq\,} 0.2260$ \cite{lui:fis:pan}. Owing to the relatively strong finite-size dependence of $\xi_{\text{L}}$ and the excessively large computational requirements near $(T_c,\rho_c)$, reliable extrapolation to $L {\,=\,}\infty$ is difficult. Nevertheless we may test for the nonanalytic behavior expected in any finite quantity \cite{fis:lan}.

\begin{figure}[ht]
\centerline{\includegraphics[width=\figurewidth]{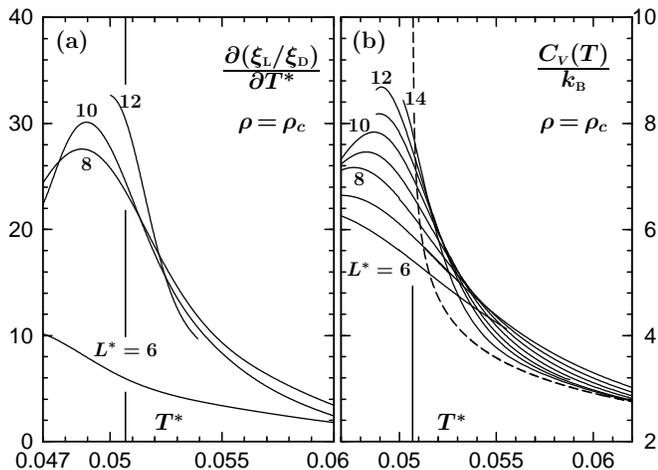}}
\caption{(a) Temperature derivative of reduced Lebowitz lengths and (b) specific heats on the critical isochore. The dashed curve approximates the bulk specific heat \cite{kim}.}
\label{fig5}
\end{figure}

On general grounds \cite{fis:lan} weak, entropy-like behavior is predicted. Thus temperature derivatives at $\rho {\,=\,} \rho_c$ should diverge like the specific heat, namely as 
\begin{equation}
  \rho C_V/k_{\text{B}} \approx A^+/t^\alpha + A^0,  \label{cv}
\end{equation}
when $t {\,=\,} (T-T_c)/T_c {\,\rightarrow\,} 0$, where $\alpha {\,\simeq\,} 0.109$ and $A^+ a^3 {\,=\,} 0.50 {\,\pm\,} 0.07$ \cite{kim} with, via a rough fit, $A^0 a^3 \simeq -0.37$. A direct comparison for finite $L$ of $\partial(\xi_{\text{L}}/\xi_{\text{D}})/\partial T$ with the specific heat is shown in Fig.~\ref{fig5} \cite{kim2}. Bearing in mind the lack of $\xi_{\text{L}}$ data near $T_c$ and its imprecision, the resemblance of the two plots is striking: we accept it as confirmation of the anticipated singularity.

\begin{figure}[ht]
\centerline{\includegraphics[width=\figurewidth]{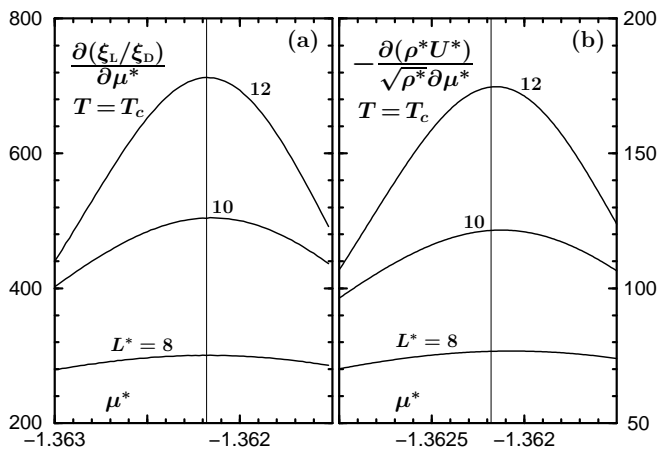}}
\caption{Derivatives on the critical isotherm of (a) the reduced Lebowitz lengths and (b) the energy densities with respect to the chemical potential $\mu^\ast$ where $\mu_c^\ast {\,\simeq\,} -1.36218$: see text.}
\label{fig6}
\end{figure}

Complementary nonanalytic behavior should arise on the {\em critical isotherm} as the reduced chemical potential $\mu^\ast {\,=\,} [\mu-\mu_0(T)]/k_{\text{B}}T$ \cite{pan:fis} varies. This is borne out by the plots in Fig.\ \ref{fig6} of $\partial(\xi_{\text{L}}/\xi_{\text{D}})/\partial\mu^\ast$ and ${\bm (}\partial(\rho^\ast U^\ast)/\partial\mu^\ast {\bm )}/\rho^{\ast k}$ with $k=\frac{1}{2}$, where $U^\ast (T,\rho)$ is the configurational energy per particle; the power $\rho^{\ast k}$ represents a convenient ``$k$-locus factor'' \cite{ork:fis:pan}. In the bulk limit both functions should, by scaling, diverge as $1/|\mu-\mu_c|^\psi$ with $\psi {\,=\,} (1-\beta)/(\beta + \gamma) {\,\simeq\,} 0.43$ \cite{lui:fis:pan,kim:fis}.

Returning to the isochore $\rho {\,=\,} \rho_c$, theory indicates
\begin{equation}
  \xi_{\text{L}}(T) = \xi_{\text{L}}^c \left[ 1 + e_\alpha t^{1-\alpha} + e_1 t + e_\theta t^{1-\alpha+\theta} + e_2t^2 + \cdots \right], \nonumber
\end{equation}
where $\theta {\,\simeq\,} 0.52$ is the leading correction exponent \cite{kim}. By making allowance for the $L$-dependence and fitting over various ranges above $T_c$ we conclude $\xi_{\text{L}}^c {\,\simeq\,} 0.30a$ and, with less confidence, $e_\alpha {\,\simeq\,} 2.6 \pm 0.2$ and $e_1 {\,\simeq\,} -2.2 \pm 0.3$.

In summary, the Lebowitz screening length, $\xi_{\text{L}}(T,\rho)$, has been studied for the restricted primitive model electrolyte via grand canonical Monte Carlo simulations of the charge fluctuations in subdomains. The corresponding area law that is asymptotically valid for large subdomains \cite{bei:fel} holds surprisingly well even in small simulation boxes, $L {\,\lesssim\,} 12a$. Finite-size effects can be understood so that the bulk, $L {\,\rightarrow\,} \infty$ limit may be extracted by extrapolation vs.\ $1/L$ for cubic subdomains and $1/L^2$ for spheres while the effective, finite-size Lebowitz lengths for slabs converge exponentially fast. Evaluation of $\xi_{\text{L}}$ for $T {\,\gtrsim\,} 10T_c$ over densities from $0.03\rho_c$ to $4\rho_c$ reveals that the {\em exact} low-density expansions \cite{bek:fis} are effective only for $\rho {\,\lesssim\,} \frac{1}{10}\rho_c$ whereas GDH theory \cite{lee:fis} reproduces well the general trends. Finally, $\xi_{\text{L}}$ remains finite {\em at} criticality but exhibits weak, entropy-like singularities on approaching $(T_c,\rho_c)$. This is the first time that charge-charge correlations and a strongly state-dependent screening length have been studied by simulations close to criticality.

National Science Foundation support via Grants CHE 99-81772 and 03-01101 (M.E.F.) and DMR 03-46914 (E.L.) is gratefully
acknowledged.

\end{document}